# Do Cosmic Backgrounds Cyclical Renew by Matter and Quanta Emissions? The Origin of Backgrounds, Space Distributions and Cyclical Earth Phenomena by Two Phase Dynamics

Eduardo del Pozo Garcia
Institute of Geophysics and Astronomy, Havana, Cuba, pozo@iga.cu, epg474@gmail.com

December 2021

**Abstract:** A numerical coincidence $\delta_p$=-0.000489 was found between the ratio of proton emission at 459-keV and cosmic background at 0.25-keV respect to proton and to electron proper energy ratio. Multiplying 459-keV and 0.25-keV by $\delta_p$, two numerical coincidences were found with 0.1-keV background and 0.06-eV relic neutrino energy, respectively.

Global criticality phenomena plus "c" and "G" discontinued decreasing in time provoke small cyclical reorganization of particles and quanta originating background emissions every 30-Myr are proposed. At one cycle beginning, electron emissions gradually originate the 0.25-keV background and, a possible 459-keV background from proton emissions lasting 0.1 Myr. At following cycle, the 459-keV quanta makes two emissions of 0.1-keV originating a diffuse background, and 0.25-keV quanta makes two of 0.06-eV originating the relic neutrinos. The $\delta_p$ is a possible physical constant that characterize the threshold, which trigger cyclical phenomena at space and Earth every 30-Myr.

This hypothesis explains: The about 30 Myr cyclic phenomena observed in tectonic, Mass Extinction, crater impact rate, and some features of space global distribution: quantized redshift, change of galaxy fractal distribution at 10 Mpc scale, galaxy average luminosity and the luminosity fluctuation of galaxy pairs are enhanced out to separations near 10 Mpc.

In addition, the traveling photons emit small part of their energy by two emissions around 282.2 GHz every 30-Myr and CMB is cyclically renew by a general global emission. The energy lost of traveling photons originate main global quantized redshift -$\delta_p$ every 30-Myr, which contributes to Hubble constant $H_0$.

Alternative #1 proposed: The 5-6 Myr intervals of features at Earth and Space are the results of the sequence of small "c" decrease at 5-6 Myr intervals ($\Delta c_{6\text{-}7Myr}/c$), which originate another sequence of small-quantized redshift. The accumulated redshift are -$\delta_p$/2 every 30-Myr. The global sequence "c" decreases every 5-6 Myr makes an accumulative difference between physical environment and matter and quanta, which internal energies correspond with a former greater value of "c". This difference has a threshold of $\Delta c/c=\delta_p/2$, at which global emissions start every 30 Myr.

The Alternative favors to explain: High redshift and Ho, high variability at less than 100 Mpc, the current abundance of light elements by the increases of nuclear emissions, because isotopes may reduce their stability.

As Universe is observed flatness in distance implies more gravitation in the past:

Alternative #2 proposed: Gravitational constant "G" also makes a sequence of discontinued decrease at 100-kyr cycle. Besides, the accumulation of several $\Delta G/G$ decreases of about $10^{-7}$ may provoke each 5-6 Myr "c" decrease, and trigger a subtle two-phase dynamics.

This Alternative favors to explain the 100-kyr problem of the ice cycle, by the increase of dust and particles at interplanetary medium, and Sun minor radiation by small expansion.

This scenario implies a subtle two-phase dynamics with 30-Myr cycle: A longer "inhibition phase" with occurrence of some biological, climate and tectonic features (BCTF) every 5-6 Myr originated by small "c" decreases, which triggered the "activation phase" of about 0.1 Myr of duration, at which particles and quanta emit, backgrounds are renewed and aforementioned Earth phenomena occur.

**Content:**

## Introduction

Several centuries were needed to solve the planetary system dynamics, although five planets are naked eye observable. The present Observational Cosmology problems are much more difficult to solve because, it is equivalent to establish the fundamental dynamics of the planetary system observing from the space-time scale of atomic nucleus.

Cosmological current extrapolations overlook the dissimilar physical behavior and phenomenology at different scales [Nottale, L., 1997]. Moreover, Statistical Physics do not need to extrapolate microcosmic properties to explain macroscopic physics. Thus, the cosmological current extrapolations to mega-scales are not well justify as right way to know the physics of the Universe large scale.

Consequently, cosmological current extrapolations to model Universe are in disadvantage with some Universe observational facts at large scales: Great Wall stability, non-gaussianity and symmetric axis of CMB and, the "missing shadows" of distant galaxies on CMB

Cosmology also assumes that redshift is only of expansive origin and, not consider the possibility that physical constants may have small changes discontinued in time at large time scale, out of our experience. In addition, other facts at near galaxies have not explanation:

- High difference between values of galaxy distance obtained by different methods, which are based on physics, and involucrate different physical constant
- Galaxy redshift high variability values at distance less than 100 Mpc, which is equivalent to analyze the Earth phenomena variability at last 100 Myr
- Gravitational lenses are not observed at Galaxy Local Group
- The observed galaxy redshift global quantization at all space directions

This paper is a "Keplerian approach": It considers that, the Universe is a "Rosetta Stone", because the Universe shows distant cosmic facts coincident in time with old Earth phenomena.

Some of these facts are on space global distributions in the scale of about 30 Mlyr:

- Galaxy average luminosity and the luminosity fluctuation of galaxy pairs are enhanced out to separations near 10 Mpc [Beisbart, C., 2000], equivalent to about 30 Myr
- Fractal dimension changes on galaxy distribution at 10 Mpc scale [Zhen, W. 1988]

By other part, coincident Earth processes with cycles of about 30 Myr are:

- The terrestrial craters [Yabushita, S., 1996]
- The terrestrial cycles of the tectonic and volcanic activities [Jain, V. E., 1984]
- The Mass Extinction of species [Raup, D. M., 1984] [Wignall, P., 2007] (Fig. 1)

Moreover, the big impacts on the Earth take place when one of the Mass Extinction is in process [Archibald, J. D., 1997] [Matsumoto, M., 1996].

Other coincident facts:

- The Local-Bubble origin about 14.5 Myr ago [Breitschwerdt, D.,2006] is time coincident with the last Mass Extinction
- A Hubble-Bubble of about 70 Mpc of radius was determined [Sinclair, B., 2010], which is time coincident with the Permic Mass Extinction
- The galaxy NGC 891 shows a high number of supernova bubbles and very hot gases. As it is about 34-40 million light-years, which is time coincident with the Mass Extinction of about 39 Million years ago at Earth

This paper considers that, due to general internal heating, comets and meteoroids increase their internal pressure and eventually make gas jet emissions perturbing their orbits, which cause, that some of them go to the inner Solar System. Therefore, the impact occurrence is increases, as well as eventually big impacts at time when a Mass Extinction is in process. The emitted materials from jets contribute to more material in space, which also increase the impact rate [del Pozo, E. 1999; --- 2000a].

The paper purpose is to know how observational facts are related one with other, and progressively deciphering the Universe dynamics in the scales of Myr and Mlyr. It tries to explain the origin of matter-background connections, and takes in considerations:

- From BOOMERanG experiment determination of Cosmic Microwave Background (CMB) power spectra follows a relatively small quantity of cosmic dark matter [Mc Gaugh, S., 2000]. The practical absence of dark matter effects in CMB is evidence that CMB is not so old. This evidence is in correspondence with the quasi-blackbody spectrum of the CMB, although the continue flow of cosmic rays, which must interact with the CMB and, in billions of years it should be deformed or eliminated. This is corroborated by the fact that, distant galaxies do not show shadows on CMB
- The global quantized redshift [Guthrie, B.N., 1997] [Tifft, W. G., 1997a] is centered in the CMB reference frame [Tifft, W.G., 1997], which implicate the existence of some relation between the CMB origin and the origin of global quantized redshift.
- The non-gaussianity of CMB was determined [Craig, J., 2004], and the existence of a Soft x-ray Diffuse Background (CXB) observed at 0.25 keV (Fig. 2), which is also emitted by the galactic halos [Barber, C. R., 1996] [Wu, K. K. S., et. al., 2006].
- The fine structure constant change may be related with some decreasing of the speed of light "c" with time [Murphy, M. T., 2000] [Barrow, J. D., 2000].

This paper put in evidence that CXB, CMB and relic neutrinos are renewed by global criticality phenomena every 30 Myr, and the $-4.89\times10^{-4}$ value is a possible physical constant that characterizes these processes.

It is propose that cyclical global change in matter is originated when an environment decrease of $\Delta c/c$ of about $10^{-4}$ is reached, which is out of our laboratory experience. Such global changes are of criticality origins provoked by the accumulation of several discontinued small decrease of light speed "c" with time, $\Delta c/c$ of about $10^{-5}$.

As the Universe is observe flat in distance and time, a sequence of several discontinued smaller decrease of gravitational constant "G", with $\Delta G/G$ of about $10^{-7}$ is proposed. This explains the 100-kyr problem [Berger, A. L., 1977] [Muller, R. A., 1994] of the ice cycle, by the increase of dust and particles at interplanetary medium [Parley, K. A., 1995], and Sun minor radiation by small expansion, as well as small Earth expansion [Witkowski, N., 1986] every 100-kyr.

Finally, the Universe large-scale structure is considers like a "Turing Structure" in three dimensions, as result of the activity of a subtle two-phase dynamics.

## Results and Discussion

### Backgrounds and Emissions

Near 459-keV several line-like features have been reported:

- Some nuclear transition processes [Creutz, E. C., 1939] [Teegarden, B. J., 2006] [Op Den Kamp, A, M, F, 1972]
- Nova process [D'Auria1, J. M., 2000] [Bishop, S., 2003]
- Gamma-ray line at galactic centre [Leventhal, M., 1980]

The ratio of 459-keV emission energy "$E_{459}$" respect to proton proper energy "$E_{0p}$":

$$E_{459}/E_{0p}= -4.89 \times 10^{-4} \quad (1)$$

And $E_{459}= (-4.89 \times 10^{-4})E_{0p}$

The diffuse background of x-ray at 0.25-keV has been established [Barber, C. R., 1996].
The ratio of 0.25-keV emission energy "$E_{0.25}$" respect to electron proper energy "$E_{0e}$":

$$E_{0.25}/E_{0e}= -4.89 \times 10^{-4} \quad (2)$$

And $E_{0.25}= (-4.89 \times 10^{-4})E_{0e}$

Then, a numerical coincidence was found, denominate "$\boldsymbol{\delta_p}$"

$$\delta_p = -4.89 \times 10^{-4} \quad (3)$$

Multiplying by $\delta_p$ to 459 keV and 0.25 keV:

$$\delta_p(459\ \text{keV})= -0.224\ \text{keV}= -2(0.112\ \text{keV}) \quad (4)$$

$$\delta_p(0.25\ \text{keV})= -0.122\ \text{eV}= -2(0.061\ \text{eV}) \quad (5)$$

They correspond with two quanta emissions of 0.1-keV diffuse background [Bloch, J. J., 1986] and two of 0.06-eV of neutrino energy [Fogli, G. L., 2007]. They are two more numerical coincidences. The first emitted 459-keV and 0.25-keV by matter at possible following processes may also emit as (4) and (5).

This put in evidence the background origins are from criticality processes cyclical in Nature, at which "$\delta_p$" characterize a threshold-value in particles and quanta emissions (Table 1).
In addition, from (1) and (2) follow:

$$E_{459}/E_{0p}= E_{0.25}/ E_{0e}$$

And $E_{459}/ E_{0.25}= E_{0p} / E_{0e}$ (6)

From (4), (5) and (6): $E_{0.1}/ E_{0.06} = E_{0p} /E_{0e}$ (7)

Then, backgrounds ratios are equal to proton electron proper energy ratio.

Moreover, the ratio of the energy of cosmic UV background maximum respect to the energy of x-ray background maximum is near $-\delta_p$ value and, also the ratio of the energy of the CMB maximum respect to the energy UV background maximum is near $-\delta_p$ value (Fig. 3) [Hauser, M. G.,2001].

Other emission-absorption process: Two narrow X-ray absorption lines in Gamma Ray Burst (GRB) were reported [Murakami, T., 1988]: 19.3 keV and 38.6 keV. The authors wrote "…why such a narrow absorption feature appears in the gamma-ray spectrum emitted by a high –temperature plasma…" However, GRB emissions are the result of radiation interaction with matter by shock scenario [Ghisellini, G. 2001].

Considering the know emission of 78 MeV of protons and meson processes, and multiplying 78 MeV by $\delta_p$, it follows:

$$\delta_p(78\ \text{MeV}) = -38.14\ \text{keV} = -2(19.07\ \text{keV})$$

Thus, more numerical coincidences with $\delta_p$ were found.

**On 30-Mlyr Space Distribution and 30-Myr Earth Cyclical Phenomena**

As we observe the Universe past, then emission at 0.25-keV from galactic halos implies a space scale of Mlyr, which correspond to Myr in time scale at Earth.

Some determinations of space global distributions show features at about 30-Mlyr: On galaxy average luminosity and the luminosity fluctuation of galaxy pairs are enhanced out to separations near 10 Mpc [Beisbart, C., 2000] and, change of fractal dimension of galaxy distribution at 10 Mpc scale [Zhen, W., 1988].

Since 30 Mlyr in distance is equivalent to 30-Myr, it is approximately the Mass Extinction cycle at Earth [Sole, R. V., 1997] [Melot, A. L., 2013], with extinction duration of about 0.1 Myr [Wignall, P., 2007] (Fig. 1). As well as the aforementioned cyclical coincident phenomena of 30-Myr, follows to consider that both groups of facts at space and Earth are originated by the same global phenomena, and might think on the following hypothesis:

*Work Hypothesis: Global criticality phenomena provoke small cyclical reorganization of particles and quanta every 30-Myr, and originate cyclical phenomena at space and Earth. At the criticality cycle, particles and quanta are compelled to emit during about 0.1 Myr. Electrons emissions originate 0.25-keV diffuse background and, proton emissions a possible 459-keV diffuse background.*

*At following 30-Myr cycle, the 459-keV quanta makes 0.1-keV emissions originating a diffuse background, and 0.25-keV quanta makes 0.06-eV emissions originating relic neutrinos. The $\delta_p$ may characterizes the threshold value at which emissions are triggered*

*May think that, a global subtle two-phase dynamics with 30 Myr cycle take place:*

*A longer "inhibition phase" with some biological, climate and tectonic features (BCTF) occurrence every 5-6Myr [Pearson, P. N. 2000], provoked by some change in physical environment, whose accumulation trigger the reorganization of particles and quanta, taking origin to the "activation phase" of about 0.1 Myr of duration including high tectonic activity and Mass Extinction [Wignall, P., 2007]. The particles and quanta emissions renewed observed backgrounds every 30 Myr cycle.*

When generalized emissions start at activation phase, some isotopes may reduce their stability and increase: the number of nuclear emissions, the quantity of light elements, and enhance the internal heating of bodies, which explain the mentioned space global distributions near 10 Mpc, and is in favors to explain:

- The current abundance of light elements
- The main spherical shape of small objects greater than 200 km in Solar System
- Major volcanic and tectonic activity [Jain, V. E. 1984]
- Comets and meteoroids make eventual jet emissions, perturbing their orbits, increasing impact frequency including big impacts at time when a Mass Extinction is in process [Archibald, J. D., 1997][Matsumoto, M., 1996]. Also, Double Mass Extinction was reported [Abbas, S.; 1998]
- Life biochemistry change and increase of genetic variability in living organisms that, originate Mass Extinction followed by a proliferation of new species, which extent about 0.1 Myr [Wignall, P., 2007] [Barrow, J. D., 1999]
- The Hubble-Bubble effects about 70 Mpc of radius [Sinclair, B., 2010] is time coincident with the Permic Mass Extinction
- The Local-Bubble origin about 14.5 Myr ago [Breitschwerdt, D., 2006] is time coincident with the last Mass Extinction

Additionally, according with the time sequence of Mass Extinction may think that a sequence of Bubbles may exist around any observer in distance and in time.

## Analyzing CMB

We observe images from long distant objects: Then, CMB quanta emissions from travelling photons as (4) and (5) must be in pairs in way that traveling photons do not lost their movement direction.

Generalizing to whole spectrum and, considering that CMB current frequency was emitted at time of last Mass Extinction. The emitter quanta frequency of CMB central frequency was:

$$\delta_p \nu_E = -2(\nu_{CMB}) \qquad (8)$$

Quanta made two emissions around 282.2 GHz (from T=2.728 K) [Fixen, D. J., 1997]:

$$\nu_E = -2(282.2)/\delta_p$$

And $$\lambda_E = -c\delta_p/2(282.2) \qquad \lambda_E = 2597\text{Å}$$

This is near the UV background maximum (Fig.3). Additionally, the ratio of the energy of the CMB maximum respect to the energy UV background maximum is near $-\delta_p$ value (Fig. 3) [Hauser, M. G.,2001].

Work Hypothesis Extension #1:

*The traveling photons around $\lambda = 2500$ Å made two emissions around CMB central frequency at time of last Mass Extinction 15 Myr ago. They made emissions every 30-Myr and CMB is cyclically renewed. The current CMB spectrum has been shape by thermalization during the last 15 Myr.*

This favors to explain the CMB quasi-blackbody spectrum, dark matter quasi-absence [Mc Gaugh, S., 2000] and the CMB "missing shadows" on distant galaxies.

As consequence, the current CMB map is the result of the propagation through the intergalactic media of the emissions around 282.2 GHz from a bubble around us of about 15 Mlyr radius.

## The Origin of Global Quantized Redshift and CMB

The traveling photon emissions $\Delta\nu = -2(\nu_{CMB})$ implies a redshift contribution every 30 Myr:

$$\Delta z = -\Delta\nu / \nu \qquad \text{using (8)} \qquad \Delta z_1 = -\delta_p \qquad (9)$$

Equivalent to a radial speed:

$$V_r = 146.6 \text{ km/s} = 2(73.3 \text{ km/s}) \qquad (10)$$

Then, another numerical coincidence was found because $V_r$ is about twice 72.1 km/s of the main global redshift quanta value to spiral galaxies in the CMB rest frame [Tifft, W.G., 1997]. This coincidence links the origins of global quantized redshift and CMB, which explains the best statistical results when redshift is measured respect to CMB rest frame [Tifft, W.G., 1997].

Work Hypothesis Extension #2:

*The energy lost of traveling photons by CMB emission originate the main global quantized redshift every 30-Myr, which contributes to the observed Hubble constant $H_0$.*

If part of traveling photons make an odd number of emissions, and lost their movement direction favors to explain:

- Galaxy diffuse light origin
- Distant objects are seen less luminous and distance measurements are overestimates [Barrow, J. D., 2000]

The interval near to 10 Mpc of greater variability in galaxy average luminosity [Beisbart, C., 2000], includes the high luminosity galaxies and low luminosity galaxies, which show high and low redshift values [Driver, S., 2004], respectively.

Thus, a possible changing factor that originates the cyclical increase of galaxy luminosity and their redshift is a sequence of small decrease of light speed "c", which compelled matter particles and quanta to emit.

Additionally, the CMB velocity may be an illusion, which imply that the galaxy activation phases are not isotropic observed. They are triggered by the propagation of a space-time perturbation. At one side we observe CMB from regions that make the last activation phase together with our galaxy, at the opposite side we observe CMB from regions, which do not reach to the last activation phase yet. Consequently, different CMB frequency is observed, and it is consider as velocity origin.

**On 6-Mlyr Space Distribution and 6-Myr Earth Cyclical Phenomena**

The quantized redshift fine structure [Tifft, W. G., 1997] may implicate global cyclical sequence of small decrease in light speed discontinued in time [Murphy, M. T., 2000].

Considering: The axis of the radio jet shape in NGC6251 shows a distinct wiggle of more than two cycles of 85 kpc, and a period of 6 Myr for an assumed speed of c/20 for the jet [Saunders, R. 1981]. That interval is coincidently with the atmospheric carbon dioxide minor values to intervals of 6 Myr (3, 9 and 15 Myr ago) [Pearson, P. N., 2000]. Thus, the two intervals are coincident.

The possible last small decrease in light speed about 2-3 Myrs ago may be related with others features at Earth:

- Central America formation by a high tectonic activity unifying North and South America
- The ancient DNA racemization shows a drastic fall near to 0 % [Poinar, H. N., 1996]
- Humans debut with abstract thinking [Tatttersall, I., 2004;2016]

*Alternative #1: Both intervals, at 5-6 Myr of features at Earth and Space are results of the sequence of small "c" decrease at 5-6 Myr intervals ($\Delta c_{6\text{-}7Myr}/c$), which originate another sequence of small-quantized redshift. The accumulated redshift is $\Delta z_2=-\delta_p/2$ every 30-Myr. The global sequence “c” decreases every 5-6 Myr makes an accumulative difference between physical environment and matter and quanta, which internal energies correspond with a former greater value of “c”. This difference has a threshold of $\Delta c/c=\delta_p/2$, at which global emissions start every 30-Myr.*

From $\mathbf{E=mc^2}$ and $\mathbf{E=h\nu}$ of quanta energy follow:

$$\Delta\nu/\nu=2\Delta c/c \qquad (11)$$

Using (9)

$$\Delta c/c=\delta_p/2 \qquad (12)$$

When the gradual emission processes in quanta start by two photon emission, the redshift:

$$\Delta z_1=-\delta_p \qquad (13)$$

Before the emission, the accumulative□ sequence of $\Delta c_{5\text{-}6Myr}/c$ reaches to $\delta_p/2$ threshold and implies $\Delta\lambda/\lambda=-\delta_p/2$

In sum every 30-Myr; the accumulated $\Delta c/c$ is $-2.44\times10^{-4}$:

$$4(\Delta c_{5\text{-}6Myr}/c)=\delta_p/2$$

Every 5-6 Myr $\quad \Delta c_{5\text{-}6Myr}/c=-6.1\times10^{-5}$.

And redshift

$$\Delta z_2=-\delta_p/2 \qquad (14)$$

The successive decrements of “c” every 5-6Myr imply successive increments of the Rydberg constant (“R”), □which imply a minor value of “R” at time when the traveling photon was emitted than the “R” current value and, may consider another contribution to redshift.

Then, two contributions to redshift occur (Fig. 4), one by “c” decrease sequence $\Delta z_2$ and, other by traveling photon emissions with a minor value of “R” $\Delta z_1$:

$$\Delta z_1+\Delta z_2=-3\delta_p/2 \qquad (15)$$

According with Table 2, the last particles background emissions took place about 15 Myr ago at Mass Extinction, after that in two times at 3 Myr and 9 Myr, the “c" value has been decrease at physical environment, which imply a current "c" value difference between environment and particles "c" internal value of about 2/5 respect to the represented Table 2 five decrements in between Mass Extinctions.

Then a laboratory experiment is possible, because current electrons and protons internal energy most have an excess, consequently may obtain emissions of 0.25 keV and 459 keV by exiting these particles approximately with 0.21 keV and 386 keV respectively.

**Quantized Redshift Contribution to Ho**

The space phenomena distributions according to Mass Extinctions time determinations [Sole, R. V., 1997] [Melot, A. L., 2013]:

- The last Mass Extinction took place 15 Myr ago, which is equivalent in distance to 15 Mlyr=4.6 Mpc
- The Mass Extinction cycle of 27 Myr, which is equivalent in distance to 27 Mlyr=8.3 Mpc intervals

Thus, the quantized redshift contribution $\mathbf{H_{QR}}$ (Table 2) to the Hubble parameter **Ho** is:

$$H_{QR}=\frac{\left[201.5+(n\text{-}1)(219.9)\right]\text{km/s}}{\left[4.6+(n\text{-}1)(8.3)\right]\text{Mpc}} \qquad n=1, 2\ldots \quad (16)$$

"n" is the number of Mass Extinction backward.

For long distance $\mathbf{H_{QR}}$=26.5 km $s^{-1}$ $Mpc^{-1}$.

The observed **Ho** from Key Project is **Ho**=78 km $s^{-1}Mpc^{-1}$ follows an expansion rate:

$$\mathbf{H_{EXP}=Ho-H_{QR}} \qquad \text{and} \qquad \mathbf{H_{EXP}} = 51.5 \text{ km s}^{-1}\text{Mpc}^{-1}$$

According Alternative#1 the two-phase dynamics scenario take place as follows (Fig. 4):
A longer "inhibition phase" of about 27 Myr with some biological, climate and tectonic features every 5-6Myr [Pearson, P.N., 2000] originated by small "c" decreases, whose accumulation triggered the "activation phase" of about 0.1 Myr of duration [Wignall, P., 2007], at which particles and quanta reduce their internal energy and emit. Thus, backgrounds are renewed and aforementioned Earth phenomena occur.

Nevertheless, the present paper may rewrite by consider that, electrons and protons make two emissions of 0.25 keV and 459 keV respectively. Thus, $\boldsymbol{\delta_p}$ value and the proposed $\mathbf{\Delta c/c}$ must be twice also. Consequently, UV traveling photons make four emissions of CMB, and the HQR value most be twice.

**Physics and its Constants**

The value intervals of the physical constants in which life is possible have been analyzed [Barrow, J. D., 1999]. Considering a smaller interval by specie life, the decrease of "c" could explain why species that survived to several Mass Extinction reach a Mass Extinction at which do not survive, when the accumulated change in a physical constant surpass one of the intervals in which specie life is possible.

However, the reported relative change of Alpha constant value [Murphy, M.T., 2000] is very small than the relative change of "c" proposed in Alternative #1. The Alpha constant not only depends of "c" value, and the change of other constant may be possible.

Seems that, Physics and its space-time constants may change discontinued in time, and different arrangements of physical constant values make stable the physical environment at different times.

These ideas conduce to incredible possibilities, for example:

- An increase of "h" value with time may maintain the Alfa value with minor changes, because the product "hc" appear in Alfa and some other formulations the change with time is not so evident.
- A more incredible increment of "¶" constant with time implies:
  - A minor Ryberg constant in the past and quantized redshift contribution
  - An increment of path length may provoke the "c" decrease with time

- The gravitational constant “G” may decrease with time according to General Relativity

By other part, the report of Double Mass Extinction processes [Abbas, S.; 1998] indicates that Alternative #1 is a simple approximation to more complex scenario. Some review analysis of former articles on Mass Extinctions [Wignal, P. B.; 2020] suggest that, they might have two main peaks of volcanism, and they determined:

“All show an initial severe extinction associated with a short-lived, but intense pulse, of anoxia followed by a recovery interval (called here the interlude phase), that lasts for ~100 ka. The second extinction pulse was associated with return of marine anoxia, which was more widespread and persisted for longer than the earlier event.”

Then, the activation phase triggering is a major change in physical environment, which is activated too. The activation phase shows at commencement a first global high activity process and a second high activity process 100,000 yr later.

*Alternative #1 Extension: The proposed emission processes at activation phase may correspond with the initial high activity and, during the following 100,000 yr physical environment is also activated and constant "c" decrease and decrease, as result the threshold value is reached again, consequently the matter particles make emission as new, and a second high activity process take place at the end of activation phase.*

Thus, contributions to redshift would be greater than (16). For example, may twice the values of redshift and $\mathbf{H_{QR}}$ showed in Table 2.

Follows:

$$H_{QR} = \frac{\left[403+(n\text{-}1)(439.8)\right]\text{km/s}}{\left[4.6+(n\text{-}1)(8.3)\right]\text{Mpc}} \qquad n=1, 2\ldots \quad (17)$$

“n” is the number of Mass Extinction backward.

For long distance $\mathbf{H_{QR}}$=53 km $s^{-1}$ $Mpc^{-1}$. This value is near Virgo Cluster **Hov**

The observed **Ho** from Key Project is **Ho**=78 km $s^{-1}Mpc^{-1}$ follows an expansion rate:

$\mathbf{H_{EXP}=Ho-H_{QR}}$ and $\mathbf{H_{EXP}}$ = 25 km $s^{-1}Mpc^{-1}$

Follows to consider that the observed redshift inside Virgo Cluster was originated by physical parameters change, because clusters of galaxies are not expanding internally, and we are observing from Virgo Cluster whose Hubble parameter is **Hov**= 56 km $s^{-1}Mpc^{-1}$ [Tamman,G. A.; 1998].

Alternatively may consider an expansion rate**:**

$\mathbf{H_{EXP}= Ho-Hov}$

Follows $\mathbf{H_{EXP}}$= 22 km $s^{-1}Mpc^{-1}$

The expansion $\mathbf{H_{EXP}}$ may be minor because, gravity implies another not-expansive contribution to redshift and minor expansion rate, which is not considered here.

Consequently, the Universe is greater and older, which could resolve the contradictions with the age of older cosmic objects [Arp, H. C., 1998] and may apply distance direct determinations methods to Ho determinations [Herrnstein, J. R., 1999].

Additionally, Alternative #1 favors to explain:

- The quantized redshifts of 18 km/s and 36 km/s (Table 2)
- Several near galaxies show high redshift and **Ho** values
- Ho show high variability at less than 100 Mpc

**The Flat Universe in Distance and Time**

As the Universe is observe flat in distance and time may think that, a time sequence of several smaller decrease of gravitational constant "G" discontinued in time, similar sequence that was proposed to "c" have been place.

Some observational fact implies more gravitation in the past, which favors to explain:

- Early Young Sun Paradox: The Earth and Mars were not ice-covered by minor distance to Sun
- Mars had a dense atmosphere and liquid water in the past
- Compact MACHOS origin: they result from early star evolution, because stars lived with less mass
- Gravitational lenses are only observed at long distance and long time ago

Additionally, may consider some facts at Earth:

- The Milankovich cycle length for Earth orbit eccentricity
- The cycle length to "100 kyr problem" in Geophysics
- The 100 kyr time interval between the two high activity at Mass Extinction activation phase [Wignal, P. B.; 2020]

*Alternative#2: Gravitational constant "G" also makes a sequence of discontinued decreases in time at 100-kyr cycle. The accumulation of several $\Delta G/G$ decreases of about $-10^{-7}$ provoke each 5-6 Myr "c" decrease, and "c" decrease accumulation trigger the two-phase dynamics.*

This alternative favors to explain:

- The 100-kyr problem of the ice cycle [Berger A. L., 1977] [Muller, R.A., 1994], by the increase of dust and particles at interplanetary medium [Parley, K. A., 1995]. Moreover, minor radiation by small Sun expansion.
- A small Earth expansion every 100-kyr [Witkowski, N., 1986]
- The Earth eccentricity cycle every100-kyr [Hay, J. D.; 1976]
- SNIa distance overestimates [Barrow, J. D., 2000], because more "G" value in the past implies smaller Chandrasekhar limit to SNIa
- Galaxy cluster dynamics with minor quantity or without dark matter.

**The Active Galaxy Duty Cycle**

A discontinue decreases in "G" may has implications to the Active Galaxy Duty cycle of about 100 Myr [Leahy, J. P., 2002] [Combes, F., 2000].

This cycle is approximately equivalent of four cycles of 27 Myr, and put in evidence the greater complexity of the cyclical change in physical environment.

Nevertheless, the active galaxy behavior may be explained as follows:

A time after of "G" decrease, galaxy nucleus might become active:

1. This change perturbs the black hole and orbit stability of surrounded material, the gravitational link of bodies, increase of dust and free objects in the galaxy. Consequently increase the material falling in black holes, which is in agreement with the x-ray observations of AGN type-I
2. Some Myr after, the free materials contribute to new star formation and provoke the increase of the star formation rate in the galaxy, which is in agreement with the infrared observation of AGN type-II.

## Further Considerations

**Physics in Different Galaxies**

Respects to CMB rest frame the main global redshift quanta are about 72.1 km/s to spiral galaxies, 36 km/s to elliptical galaxies and 24 km/s to quasars [Tifft, W. G., 1997]. Generalizing using (8), (9) and (11) follows:

$$\Delta z=-\delta_p/k \qquad k(\Delta\nu/\nu)=\delta_p \qquad k(\Delta\nu/\nu)=2\Delta c/c$$

Where: k=1 to spirals, k=2 to ellipticals and k=3 to quasars

Background emission processes with different threshold implies different physical environment in different galaxies, a challenge to Physics, but the so active collision between a spiral galaxy and a quasar is also a challenge to understand. Different physical environment may also be related with the intrinsic redshift in galaxies and different redshift at galaxy-pairs [Arp, H.C, 1998] [Bell, M. B., (2003)].
These ideas conduce to incredible galaxy types generalization $\mathbf{E_k=mc^{2k}}$

**An Alternative Approach to Cosmology**

The processes of cyclical emissions every 27 Myr imply a sequence of emissions where the former emitted quanta are compelled to emit at the following cycle, and the energy of the emissions decrease and decrease to generate Extremely Large Waves (ELW), which may reach long waves of hundreds and thousands of Mlyr of wavelength.

Elementary particles conforms the neutron star, which are stably macrocosmic objects. Also may think that the ELWs at some threshold value may conform stably megacosmic objects, as ELWs megaclouds. The ELWs megaclouds may constitute dense energetic environment in increase, that may be consider as the “dark energy” environment, which may explain the particles generation in vacuum experiments.

The ELWs megaclouds in collision excite one megacloud with another; and may trigger processes of baryonic matter generation taking origin to Super Massive Black Holes (SMBH), the galaxy-forming processes and the Galaxy Clusters (GCs). These ideas may explain the filament shape distribution of galaxy super clusters.

Additionally, according to the present paper the current CMB is the result of global emissions that took place about 14-15 Myr ago, and may think that the symmetry axis of the CMB is the result of two ELWs megaclouds in collision, which took origin to galaxies of Virgo cluster.

Considering the present paper proposition of more value to light speed in the past implies that also gravitation was greater because the Universe is observe flat.

Besides, the distance determinations by SNIa imply an accelerated expansion in the past, but direct measurements of distances [Herrnstein, J. R., 1999] show about 25 - 30% minor distance values to near galaxies M100 and M106.

Nevertheless, more "G" value in the past implies that Chandrasekhar limit to SNIa was smaller, and distant supernovas SNIa explode with minor mass and show less luminosity. Thus, the SNIa distance determinations are overestimates. Consequently, a correction to the SNIa distance determination by more "G" value in the past is required.

Furthermore, at great distance, not only primordial galaxies are observed. Groups of big galaxies also are reports, which imply an older Universe.

The Alternative#1 of present paper is a simple approximation to more complex scenario, which implies that a part of observe redshift is not expansion. Consequently, may think that the non-expansive component of observe **Ho,** the quantized redshift contribution $\mathbf{H_{QR}}$ is greater than values obtained by relations (16) and (17).

Moreover, more gravitation in the past implies that the gravitational redshift reduces the estimated $\mathbf{H_{EXP}}$ expansion value. Then, a value around $\mathbf{H_{EXP}}$ =7 km $s^{-1}Mpc^{-1}$ may consider.

According to the former ideas may postulate:

1. "The Universe has the property of maintain the relation between light-speed and gravitation in way that it conserves flat"
2. “The ELWs megaclouds constitute the physical environment of vacuum, which is a pre-baryonic environment and may be are the result of a mega-quantification [Nottale, L., 1997]”
3. "The ELWs megaclouds are intrinsically expansive and may fit the Friedman solution to Einstein General Relativity with cero baryonic density"
4. "Two ELWs megaclouds in a collision process excite one cloud with other, consequently a sequence in time of mini big bangs take place, which generates baryonic matter, similarly as reports of particle generation in vacuum experiments. Thus, excited ELWs megaclouds in collision take origin to a sequence of supermassive black holes and galaxy formations.”

5. “The galaxies formation processes increase the Universe mass and diminished the ELWs density and gravity necessary to Universe self-regulation. Consequently the Universe self-organize to critical density $\Omega = 1$ and the current expansion is about $\mathbf{H_{EXP}} \approx 7$ km $s^{-1}Mpc^{-1}$"

Then, through the difference in distance determinations between 7 km $s^{-1}Mpc^{-1}$ and distance determinations by SNIa may determinate the behavior of "G" constant in distance and time.

The Cycle of Matter and Energy:

The baryonic matter through the two phase dynamics contribute to ELWs megaclouds formations, and ELWs megaclouds collisions generate baryonic matter and galaxy formations, taking origin of a cycle of matter and energy.

Then, follows a question: What were first the ELWs megaclouds or the baryonic matter?

Before a Baryonic Big Bang: May think in a first overwhelming megacloud that grows and grows, it lost stability and it divided in two or more parts, take origin to gravitation and start the megaclouds collisions and the proposed “Cycle of Matter and Energy”. According to the former ideas, a better name to “dark energy” is “Creation Energy”.

**Conclusions**

The current aforementioned backgrounds, and possible background at 459-keV are cyclically renewed by criticality processes, where -4.89 10-4 is the threshold value.

A two-phase dynamics originated by time discontinued decrease of "G" and "c" trigger global emissions of internal reorganization of matter and quanta, and cause the change in fractal dimension of distribution of galaxies at 10 Mpc scale.

As a consequence the distribution of galaxies seems to snail picture, it seems as 3D Turing-structure deformed by different time-delay with distance.

The report of alpha variation and discontinued decrease of "c" and "G" may imply that other physical constants might also change as the result of global space-time reorganization at each activation phase. Remain some open question:

- Why constants most change with time?
- Is the two-phase dynamics a space-time process looking for Universe stability?
- How many activity phases remain to reach at Universe stable state?
- Will be an activation phase at which particles reach a threshold and lost stability

**Acknowledgement**

For motivation in this thematic: Michael J. Benton in “Interpretations of Mass Extinction”, Nature vol. 314, 1985.
For discussions: Mario Campos, Giraldo Alayón, Rina Pedrol, Roman Duarte, Jorge Miguel Garcia, Hugo Perez, Efren Jaimez, Angel Augier, Mario Naito, José Alvarez Lorenzana, Francisco Gonzalez Veitia and José Rodriguez Caballero.
For send papers: John D. Barrow, William Tifft, Paul Wignal, Laurent Nottale, W. M. Napier, Stacy Mc Gaugh and Jaint Narlikar.
For attention and WEB published to Arxive of Cornell University, USA

| **Table 1: Background Emissions** | | |
|---|---|---|
| **Proper Energy $E_0$** | **Background $(-4.89\times10^{-4})\ E_0$** | **Background $(-4.89\times10^{-4})^2\ E_0$** |
| Electron **$E_{0e}$** one emission | X-ray 0.25 keV two quanta emissions | 0.061 eV (neutrino) |
| Proton **$E_{0p}$** one emission (proposed) | $\gamma$-ray 459 keV two quanta emissions | 0.112 keV X-ray |

| Table 2: Scenario with Distance and Time of Quantized Redshift Contribution to $H_0$ | | | | |
|---|---|---|---|---|
| **Sequence of Δc in Distance (backward in time)** | **Distance Mlyr (time-Myr)** | **Accumulated Redshift (km/s)** | **Effects at Earth** | **$H_{QR}$ km $s^{-1}$ $Mpc^{-1}$** |
| Δc | 2.5-3 | 18.3 | BCTF* | 18 |
| 2Δc | 8-9 | 18.3 | BCTF | 9 |
| 3Δc Matter reorganization: Traveling photons redshifted by emissions | $M_1$=14 | Photon redshifted 36.6+18.3+146.6 = 201.5 | Mass Extinction | 46 |
| 4Δc | $M_1$ +5.2 | 201.5+18.3= 219.8 | BCTF | 37 |
| 5Δc | 24.4 | 238 | BCTF | 32 |
| 6Δc | 29.6 | 256 | BCTF | 28 |
| 7Δc | 34.8 | 275 | BCTF | 26 |
| 8Δc Matter reorganization: traveling photon redshifted by emissions | $M_2$= 40 | Photon redshifted 275+18.3+146.6= 439 | Mass Extinction | 36 |
| 9Δc | $M_2$ +5.2 | 457 | BCTF | 34 |
| 10Δc | 50.4 | 475 | BCTF | 31 |
| 11Δc | 55.6 | 493 | BCTF | 29 |
| 12Δc | 60.8 | 512 | BCTF | 27 |
| 13Δc Matter reorganization traveling photon redshifted by emissions | $M_3$= 66 | Photon redshifted 475+18.3+146.6 = 658 | K-T Mass Extinction | 33 |

Table 2: An illustration of quantized redshift contribution to H0 from the three last Mass Extinction, which took place about 14-15 Myr, 39-40 Myr and 65-66 Myr ago is presented. First, it is presented the last "c" constant decrease at 2.5-3 Myr and 8-9 Myr ago after the last Mass Extinction 14 Myr ago, followed by the decrease of "c" constant at intervals of about 5.2 Myr in between Mass Extinctions, in way that each Mass Extinction took place at the fifth "c" constant decrease. The 5.2 interval length corresponds to the 26 Myr interval in between Mass Extinctions.

*(BCTF) represent the occurrence of some biological, climate and tectonic features

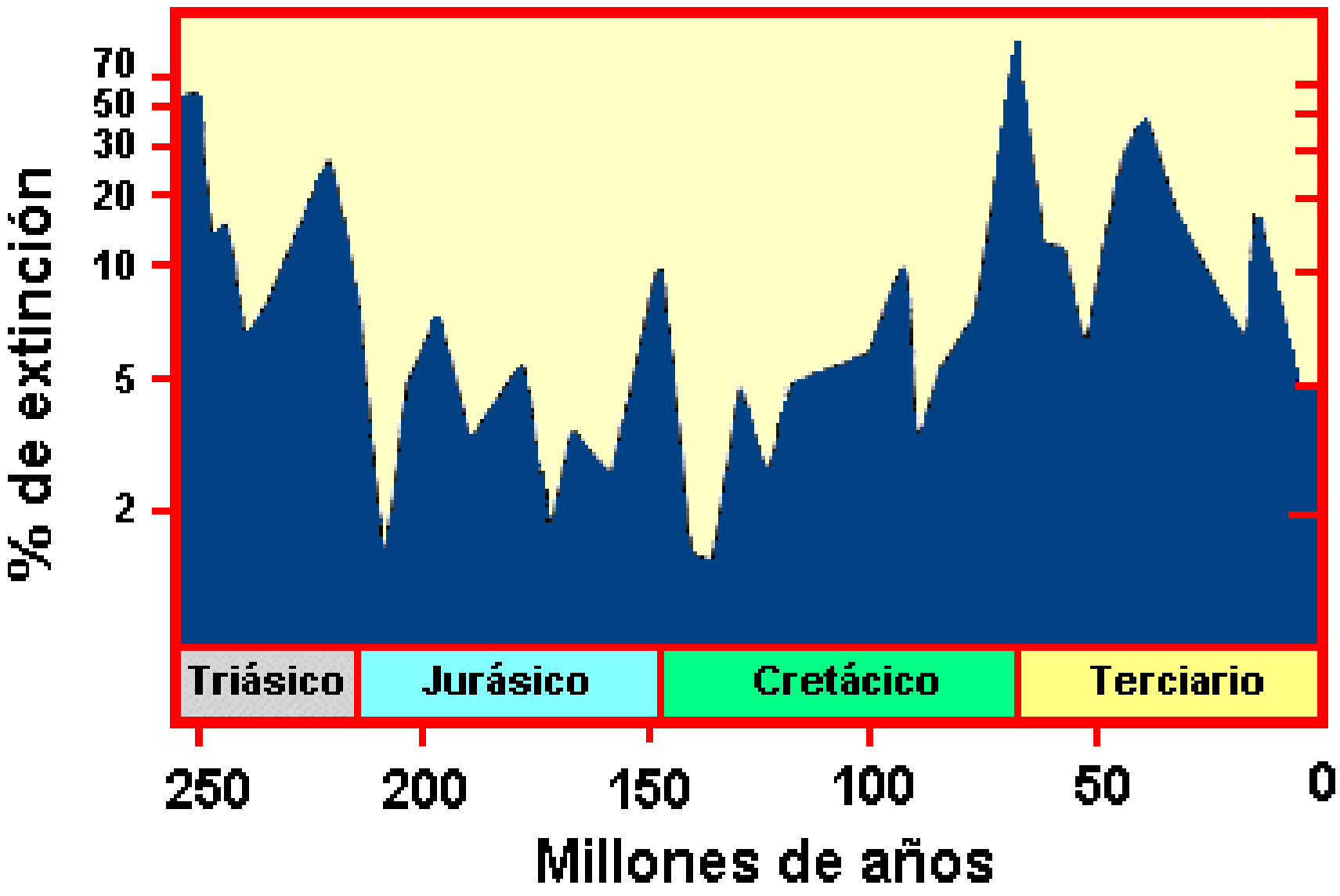

**Fig. 1:** Extinction percent during the last 250 Myr

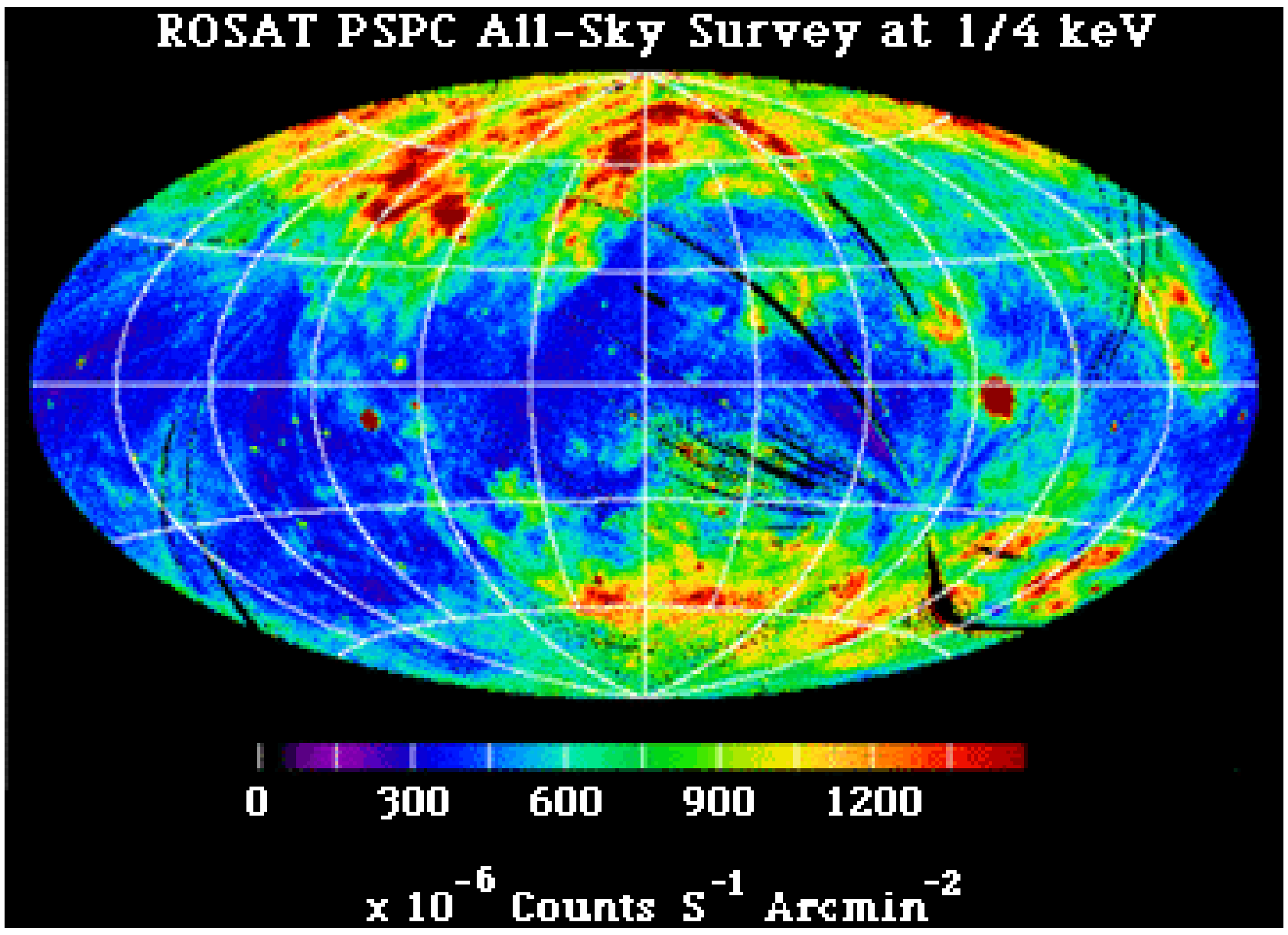

**Fig. 2:** Soft X-ray Diffuse Background: ROSAT 0.25 keV All-Sky Map

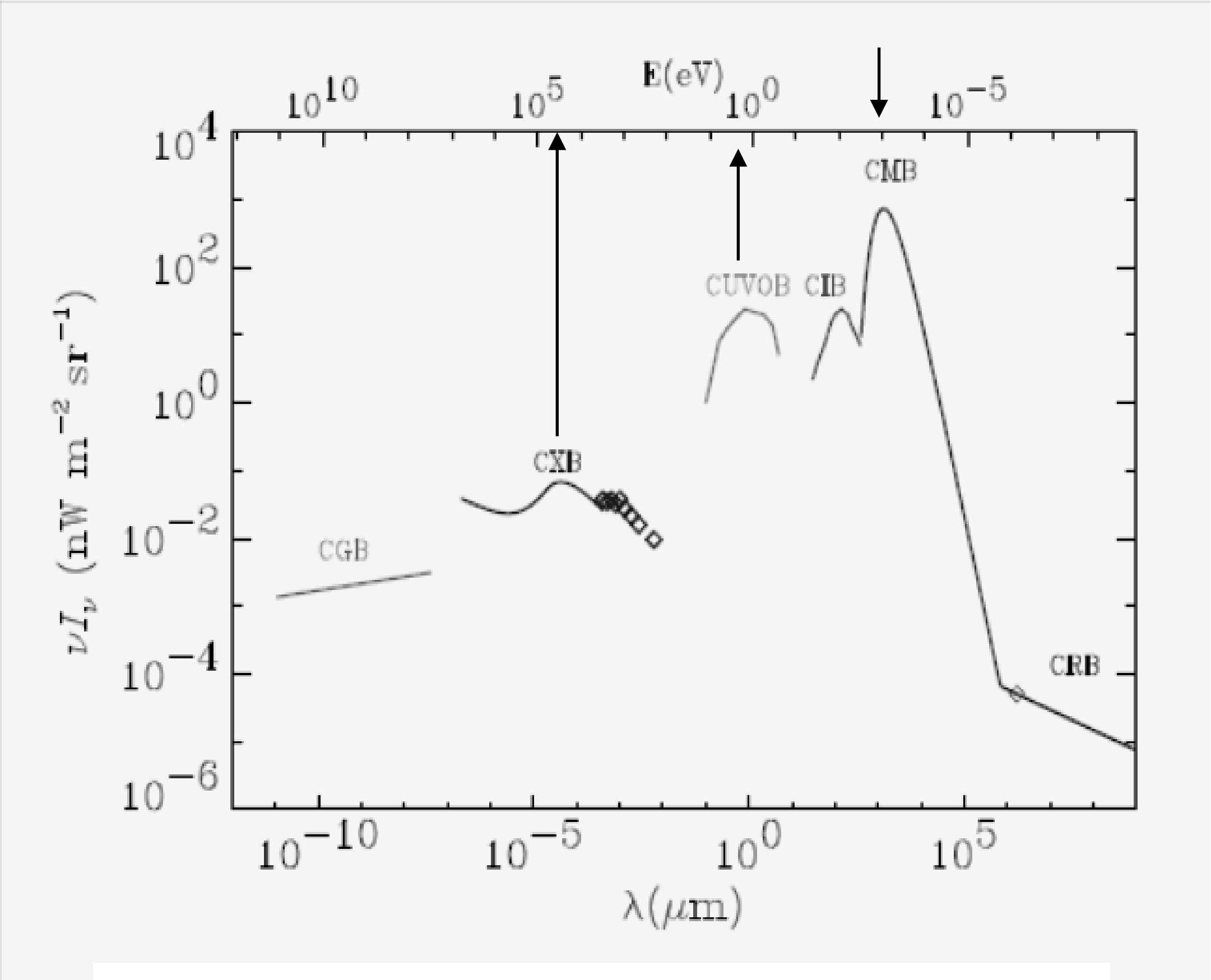

**Fig. 3:** Spectral energy of various cosmic backgrounds: CGB, CXB, CUVOB, CIB, CRB and CMB

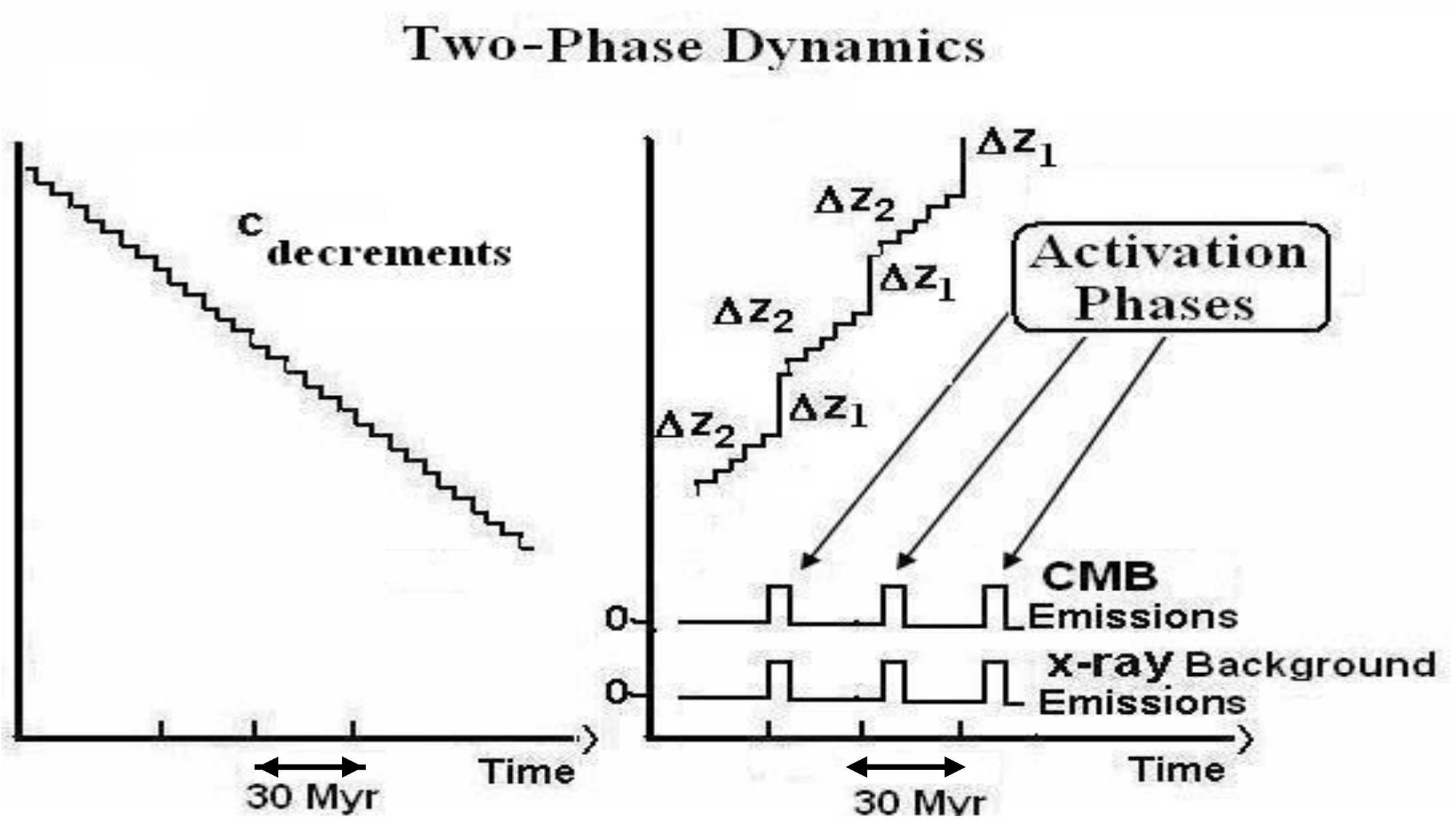

**Fig. 4:** The scenario of the cyclical behaviour provoked by successive “c” decrements (at left). As a consequence the redshift z shows pattern behaviour (at right) with several small $\Delta z_2$ increments (quantized redshift fine structure). After four $\Delta z_2$ follows a greater one $\Delta z_1$ increment, which take place when matter particles self reorganized their internal energy to small “c” value every 30 Myr. Gradual emissions from UV of CMB and X-rays emissions from electrons renew some observed backgrounds. And possible background at 459 keV from protons and other radiations may provoke Mass Extinction and proliferation of new species. The y-axe has arbitrary scales.